# Combined frequency comb and continuous wave cavity-enhanced optical-optical double-resonance spectrometer in the 1.7 µm range

Vinicius Silva De Oliveira,[1] Adrian Hjältén,[1] Isak Silander,[1] Andrea Rosina,[1] Michael Rey[2], Kevin K. Lehmann,[3] and Aleksandra Foltynowicz[1,*]

[1]*Department of Physics, Umeå University, 901 87 Umeå, Sweden*
[2]*Groupe de Spectrométrie Moléculaire et Atmosphérique, UMR CNRS 7331, BP 1039, F-51687 Reims Cedex 2, France*
[3]*Departments of Chemistry & Physics, University of Virginia, Charlottesville, VA 22904, USA*
*\*aleksandra.foltynowicz@umu.se*

**Abstract:** We present an optical-optical double-resonance (OODR) spectrometer based on a 3.3 µm continuous wave pump and two cavity-enhanced probes: a frequency comb tunable in the 1.64 – 1.8 µm range, and a comb-referenced continuous wave (CW) laser tunable in the 1.6 – 1.75 µm range. The comb probe provides broad spectral coverage (bandwidth up to 7 THz) for simultaneous detection of many sub-Doppler OODR transitions with sub-MHz line position accuracy, while the CW probe allows targeting individual transitions with kHz accuracy and higher signal-to-noise ratio in shorter time. Using the pump stabilized to the frequency of the R(0) transition in the $v_3$ band of methane and the comb probe covering the 5550 to 6070 cm$^{-1}$ interval, we detect 37 ladder-type transitions in the $3v_3 \leftarrow v_3$ band region and 6 V-type transitions in the $2v_3$ band region and assign them using available theoretical predictions. Using the CW probe, we measure selected ladder- and V-type transitions with much higher precision. We also detect Lamb dips in the R(0) – R(3) transitions of the $2v_3$ band and report their center frequencies with kHz level accuracy. The synergy effects of the comb- and CW-OODR open new possibilities in precision spectroscopy of levels that cannot be reached from the ground state.

## 1. Introduction

Optical-optical double-resonance (OODR) spectroscopy of static gas samples is a well-established technique for investigation of excited molecular states [1], their dynamics [2] and collisional effects [3], as well as for trace gas detection [4]. It allows measurements of transitions from/to selectively populated molecular levels and reaching levels that are not accessible from the ground state. In the early days, the technique was implemented using tunable pulsed (ns) lasers whose high power and broad spectral coverage enable efficient pumping and surveys of broad spectral ranges. However, the linewidths of those lasers are of the order of or larger than the Doppler widths of molecular transitions, which limits the spectral resolution [5]. More recently, OODR spectroscopy has been implemented using narrowband continuous wave (CW) lasers that allow measurements with sub-Doppler resolution. A CW pump laser excites a selected velocity group, which makes the probe signals free of Doppler broadening. Three types of collision-free probe transitions are possible in OODR: ladder-type, which start from the upper level of the pump transition and reach higher energy levels, and appear as sub-Doppler absorption lines; V-type, which share the lower level with the pump transition and appear as sub-Doppler dips in the corresponding Doppler-broadened transitions; and Λ-type, which share the upper state with the pump transition and appear as sub-Doppler lines of optical gain. The Λ-type transitions appear only if the pump frequency is higher than the probe frequencies. Thus, in our work, where the pump frequency is lower than the probe frequencies, only ladder- and V-type transitions appear, as illustrated in Fig. 1.

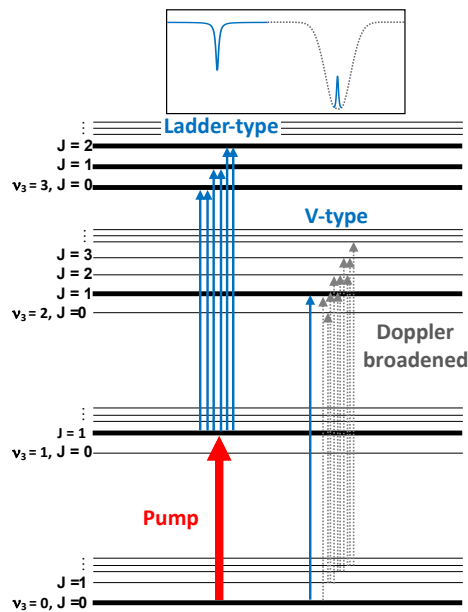

Fig. 1. Simplified schematics of energy levels of methane addressed using MIR-NIR OODR spectroscopy in this work. The pump transition, R(0) in the $\nu_3$ band, is indicated with the thick red arrow. Dotted grey arrows indicate the Doppler-broadened probe transitions from the thermally populated levels in the ground state to the $2\nu_3$ states. The sub-Doppler OODR transitions are indicated with solid blue arrows; absorption-like ladder-type transitions occur between the level populated by the pump and upper states in the $3\nu_3$ range with $J = 0$ to 2, while the V-type transitions occur as dips in the Doppler-broadened transitions sharing the lower level ($J = 0$) with the pump transition. Note that the final states reached by the V-type and ladder-type transitions involve a combination of vibrational levels in the P = 4 and 6 polyads, respectively, that are not shown for simplicity.

Various combinations of near-infrared (NIR) and mid-infrared (MIR) CW lasers have been used for OODR. Karhu *et al.* [6] used a 3 μm CW optical parametric oscillator (OPO) as a pump and external cavity diode laser (ECDL) tunable between 1520-1570 nm as a probe to measure 10 sub-Doppler ladder-type transitions reaching symmetric vibrational states of acetylene in the 10000 cm⁻¹ range and determine band parameters. The probe was cavity-enhanced and the spectra were measured using cavity ringdown spectroscopy (CRDS), with frequency accuracy limited by the wavemeter and the long-term linewidth of the ECDL probe. The accuracy of this MIR-NIR OODR spectrometer was vastly improved by referencing the pump and probe lasers to a frequency comb [7], which allowed determination of the position of the R(17) line in the $\nu_1 + 2\nu_3 \leftarrow \nu_3$ band of acetylene with 13 kHz uncertainty. More recently, Tan *et al.* [8] used a comb-referenced 4.3 μm CW-OPO as pump and a comb-referenced cavity-enhanced 1.5 μm ECDL as probe with CRDS detection to measure the P(15) transition in the (00041)-(00011) band of ¹³CO₂ with 4.3 kHz accuracy. NIR-NIR OODR has been demonstrated by Hu *et al.* [9, 10], where the pump and probe were two comb-referenced 1.5 μm ECDLs resonating in an enhancement cavity and detected by CRDS. This comb-locked cavity-assisted DR (COCA-DR) spectrometer was used to measure transitions to highly excited levels in CO₂ [9] and CO [10], and to measure the frequency of the R(1) (2-0) transition of HD, using the V-type excitation [11], all with sub-MHz linewidths. Finally, MIR-MIR OODR has been implemented by Okubo *et al.* [12] in an absorption cell (*i.e.*, with no cavity enhancement) using two comb-referenced difference-frequency-generation sources to measure 9 transition in the $2\nu_3 \leftarrow \nu_3$ range of methane with 10 kHz accuracy. Jiang and McCartt [4, 13] implemented cavity-enhanced MIR-MIR OODR, which they termed two-color intracavity pump-probe CRDS, for highly selective trace gas detection. Using two 4.5 μm quantum cascade lasers locked to a high-finesse cavity, they measured ladder-type transitions in $N_2O$ [13] and later used the technique for detection of $^{14}CO_2$ with ppq sensitivity [4]. Tan *et al*. also suggested the use of the COCA-DR method for $^{14}CO_2$ detection at sub-ppt levels [14], but the experimental realization remains to be seen. It should be noted that when pump and probe are both resonant with the enhancement cavity, they need to be tuned simultaneously, which adds a level of complexity to the retrieved signals, as discussed in Refs. [9, 13].

CW-OODR spectroscopy allows measurements with sub-Doppler resolution, and – when referenced to frequency combs – with kHz level accuracy of line positions. However, its spectral coverage is limited by the tunability and required scanning time of the CW lasers. Much wider simultaneous spectral coverage, combined with sub-Doppler resolution, is provided by optical frequency combs. Frequency combs were first used for OODR of Rb, an atomic species that is easy to saturate [15, 16]. Recently, our group demonstrated comb-based OODR of a molecular species with unprecedented combination of sub-Doppler spectral resolution and wide spectral coverage [17]. Using a 3.3 μm CW-OPO as pump and a frequency comb probe centered around 1.67 μm, we measured ladder-type transitions in the $3\nu_3 \leftarrow \nu_3$ range of methane and V-type transitions in the $2\nu_3$ range [18]. In this first demonstration, the methane sample was held in a $LN_2$-cooled single-pass cell, which limited the sensitivity and frequency accuracy of the method. We later introduced an enhancement cavity for the comb probe [19], which allowed detection of a much larger number of probe transitions with better signal-to-noise ratio (SNR) and ~150 kHz frequency precision. Most recently we used this method to measure and assign hot-band methane transitions from excited rotational levels [20].

Here we present a combined comb- and CW-laser-based MIR-NIR OODR spectrometer that allows broadband surveys of wide spectral ranges around 1.7 μm with sub-MHz frequency accuracy using the comb probe, and then targeting individual lines with much higher precision, accuracy and SNR using the comb-referenced CW probe. We report new ladder-type and V-type transitions of $CH_4$ measured with this spectrometer with pump stabilized to the frequency of the $\nu_3$ R(0) transition.

Compared to our previous cavity-enhanced comb-based OODR spectrometer [19, 20] we extended the spectral coverage of the probe beyond 1.7 μm and introduced a new frequency referencing scheme for the pump. Previously, the pump was stabilized to the center of the selected pump transition using a Lamb-dip lock in a reference cell. The disadvantage of this scheme is that it allows stabilizing the pump only to strong transitions that can be saturated in the reference cell, and the frequency of the pump is then known with the accuracy of previous measurements available in the literature and databases. Here, we reference the CW-OPO to a MIR optical frequency comb, which allows stabilizing it at an arbitrary offset from the pump transition and provides absolute frequency readout. We use both locks and compare their performance.

Compared to previous MIR-NIR CW-OODR demonstrations [7, 8], in our spectrometer the pump and probe lasers are referenced to two separate frequency combs instead of one, which allows fully independent tuning of both laser frequencies. The spectrometer operates with a CW probe tunable in the $1.6 - 1.75$ μm range, suited for methane, while the previous ones probed in the telecom range. We use the spectrometer to detect ladder-type, V-type and Lamb-dip transitions, with frequency accuracy on the single kHz level. For measurement of OODR probe transitions, instead of CRDS we employ lock-in detection by amplitude modulating the pump power, yielding a better level of absorption sensitivity despite the use of a lower finesse optical cavity.

We stabilize the pump to the R(0) transition in the $\nu_3$ band, as shown in Fig. 1. Using comb-OODR we detect 37 ladder-type transitions reaching final states in the $8550 - 9100$ cm$^{-1}$ range, of which 32 are new, and 6 V-type transitions (5 new). We compare the ladder-type transition frequencies and intensities to predictions from a new effective Hamiltonian [21], ExoMol [22, 23] and TheoReTS/HITEMP [24], demonstrating again [20, 25] the higher accuracy of the new Hamiltonian predictions. Using CW-OODR we detect the Lamb-dip and the V-type feature in the $2\nu_3$ R(0) transition and determine the frequency accuracy by comparison to previous results of Votava *et al.* [26]. We then report improved center frequencies for three R(3) transitions in the $2\nu_3$ band. We also use the CW-OODR to measure a selected ladder-type transition to determine its assignment from polarization dependent intensity ratios [19, 27].

## 2.  Experimental setup and procedures

The experimental setup, shown in Fig. 2, is based on a high-power 3.3 μm CW-OPO acting as a pump, and two alternately employed cavity-enhanced probes operating around 1.7 μm: a 250 MHz comb probe (Fig. 2a) and a CW probe (Fig. 2b). The frequency of the pump is stabilized either using a Lamb-dip lock, or by referencing to a MIR frequency comb, as described in Section 2.1. The probe light is led to the cavity via a fiber-coupled optical circulator (OC), which allows convenient switching between the comb and CW probe sources by reconnecting optical fibers leading to the OC. At the output of the OC fiber, the slightly elliptical polarization (around 3% along the minor axis) is cleaned using a polarizer. The pump power is controlled using a half-wave plate and a polarizer, and its polarization with respect to that of the probe is set using a half-wave plate after the polarizer. The pump and probe beams are combined in front of the cavity and then separated behind the cavity using dichroic mirrors (DM). The power of the pump beam transmitted through the DM after the cavity is monitored using a thermal power meter. The probe beam reflected from that DM is directed to the detection system: a Fourier transform spectrometer (FTS) (Section 2.2) for the comb probe, or a photodiode (PD4) for the CW probe (Section 2.3).

The sample cavity, which is the same as in Ref. [20], is resonant for the probe and single pass for the pump. The cavity length is 60 cm, so that its free spectral range is matched to the comb probe repetition rate, $f_{rep}$, of 250 MHz. The cavity mirrors (Layertec) have YAG substrates with 1 m radius of curvature. The mirrors have reflectivity of ~99.7% in the 5400-6100 cm$^{-1}$ range and antireflection coating in the 2700-3100 cm$^{-1}$ range, resulting in 95% cavity transmission for the pump. The cavity spacer is made of stainless steel, and one of the mirrors is mounted on a piezoelectric transducer (PZT) that allows controlling the cavity length. The sample cavity is connected to a supply of CH$_4$ (Air Liquide, 99.995% purity), a capacitance manometer to measure sample pressure, and a turbomolecular pump.

The probe beam is mode-matched to the TEM$_{00}$ cavity mode, with a beam radius of 0.5 mm at the waist, using a telescope (not shown). The same telescope is used for mode-matching the comb and the CW probe beam. The pump beam is focused at the center of the cavity with radius at focus of 0.7 mm, so that its Rayleigh range of 45.8 mm is equal to the probe's to maximize their spatial overlap. The comb and CW probe lasers are locked to the cavity using the Pound-Drever-Hall (PDH) method [28], and absolutely stabilized as described in Sections 2.2.1 and 2.3.1, respectively. All fixed and tunable frequency sources and counters are referenced to a GPS-disciplined Rb clock with 10$^{-12}$ relative stability at 1 s (Symmetricom, TSC4410).

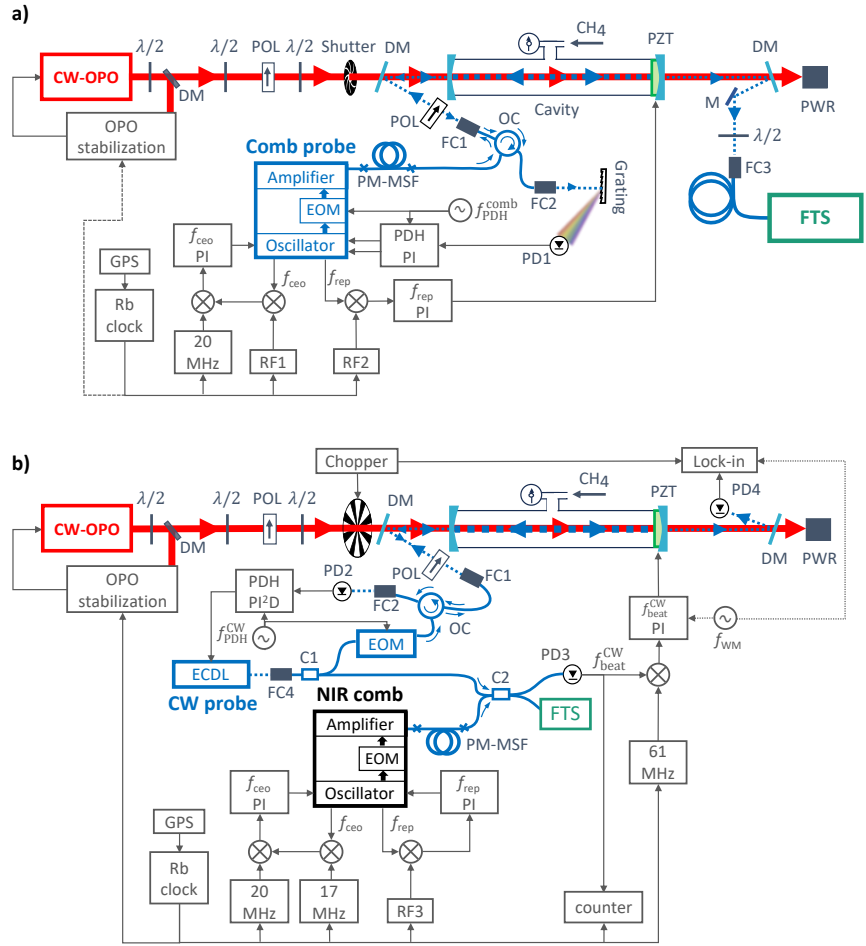

Fig. 2. Experimental setup with a) the comb probe and b) the CW probe. CW-OPO: continuous wave optical parametric oscillator that is used as pump. λ/2: half-wave plate. DM: dichroic mirror for combining/splitting the pump and probe beams. PZT: piezoelectric transducer. POL: Polarizer. PD1 to PD4: photodiodes. PWR: power meter. FC1 to FC4: fiber collimators. C1 and C2: fiber combiners. OC: fiber optical circulator. EOM: electro-optic modulator. PM-MSF: polarization-maintaining microstructured silica fiber. In b), the lock-in amplifier is used for WMS Lamb dip measurements (dashed line circuit) and for OODR measurements using a chopper for the pump beam (solid line circuit).

## 2.1 Pump

The pump is the idler of a singly-resonant continuous-wave optical parametric oscillator (CW-OPO, TOPTICA, TOPO) seeded by a narrow-linewidth 1 μm ECDL (TOPTICA, CTL PRO, specified short-term linewidth <10 kHz at 5 μs). The frequency of the idler can be stabilized to the center of the $\nu_3$ R(0) transition in two ways: using a Lamb-dip lock in a reference cell or by direct referencing to a MIR frequency comb, as shown in Fig. 3 and described below. A small fraction of the idler power (20-40 mW) is reflected using a combination of a half-wave plate and a dichroic mirror (used for its polarization sensitivity) and directed on a pellicle beamsplitter. The beam transmitted through the beamsplitter is coupled to the Lamb-dip reference cell, while the reflected beam is spatially overlapped with the beam of the MIR comb. Thus, the Lamb-dip signal and the beat note with the MIR comb can be observed simultaneously.

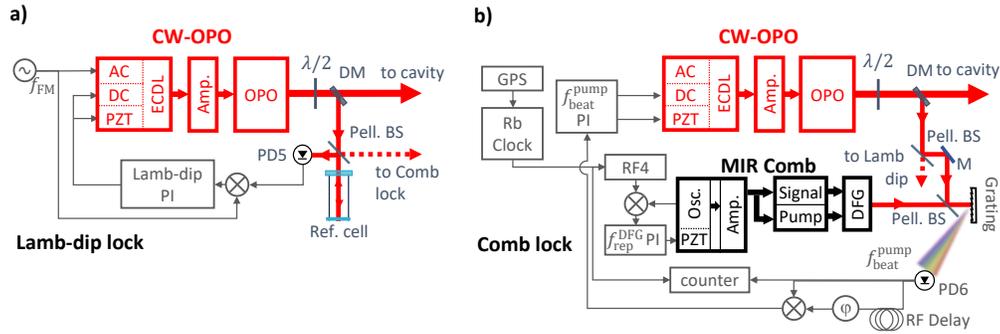

Fig. 3. Schematics of the CW-OPO frequency stabilization schemes. a) Lamb-dip lock, and b) MIR comb lock. AC: AC-coupled seed laser current control. DC: DC-coupled seed laser current control. PZT: piezoelectric control of the ECDL seed wavelength. Amp.: seed amplifier. $f_{FM}$: modulation frequency. $\lambda/2$: half-wave plate. DM: dichroic mirror. Pell. BS: pellicle beamsplitter. Ref. cell.: reference cell. PD5 and PD6: photodiodes. RF Delay: delay line for the offset frequency lock. φ: phase shifter.

### 2.1.1 Lamb-dip lock

The Lamb-dip lock, shown in Fig. 3a), is the same as in Ref. [20]. Shortly, the Lamb dip is detected using frequency modulation (FM) spectroscopy in a 30-cm-long reference cell containing $50 – 60$ mTorr of pure $CH_4$. The cell has a $CF_2$ window at the input and a flat silver mirror at the back to reflect the beam, which is then picked off by the pellicle beamsplitter and sent to a fast TEC-cooled HgCdTe detector (PD5, VIGO Photonics, PVI-2TE-5). The AC input of the seed laser current is modulated at $f_{FM} = 27$ MHz, which results in phase modulation of the idler. The detector signal from PD5 is demodulated at $f_{FM}$ to generate an error signal that is sent to a proportional-integral servo controller (Lamb-dip PI, New Focus, LB1005). Correction is applied in parallel to the DC-coupled current input of the seed laser for fast feedback, and to the PZT input for slow feedback.

### 2.1.2 MIR optical frequency comb lock

Alternatively, as shown in Fig. 3b), the CW-OPO idler is referenced to a 125 MHz MIR optical frequency comb produced by difference frequency generation (DFG) from a single oscillator [29]. The MIR comb is inherently $f_{ceo}$-free, and it is stabilized by locking the 4th harmonic of the repetition rate, $f_{rep}^{DFG}$, to an RF source (RF4, Analog Devices, AD9914), as described in Ref. [30]. The Allan deviation of the 4th harmonic of $f_{rep}^{DFG}$ at 1.1 s integration time is 3 mHz, corresponding to 540 Hz frequency jitter for the mid-IR comb teeth. The MIR comb beam is collinearly combined with a fraction of the CW-OPO idler beam using another pellicle beamsplitter and dispersed using a grating. The dispersed light is incident on a fast TEC-cooled HgCdTe detector (PD6, VIGO Photonics, PVI-4TE-4) to detect the beat signal $f_{beat}^{pump}$. We use an offset lock scheme similar to Ref. [31] with a 20-m delay line to stabilize the beat frequency at around 22 MHz. The error signal is sent to PI loop filters ($f_{beat}^{pump}$ PI, New Focus, LB1005) and the correction signals are sent to the PZT and the DC coupled current input of the CW-OPO seed laser. Rb-referenced frequency counters are used to monitor $f_{rep}^{DFG}$ (Tektronix, FCA3003, 1.1 s gate time, not shown in Fig. 3b) and $f_{beat}^{pump}$ (Rohde & Schwarz, HM8123, 0.7 s gate time), and the pump frequency is calculated as $\nu_{pump} = n f_{rep}^{DFG} \pm f_{beat}^{pump}$. The free-running frequency drift of the CW-OPO idler is around 1 MHz/s and stepping $f_{rep}^{DFG}$ allows identification of the beat note sign.

Before locking the CW-OPO idler to the comb, we tune its frequency to the center of the $\nu_3$ R(0) transition by observing the DC Lamb-dip signal from the reference cell. Using the transition frequency measured by Abe *et al.* [32], and the counted $f_{\text{beat}}^{\text{pump}}$ and $f_{\text{rep}}^{\text{DFG}}$, we calculate the comb mode number, $n$, closest to the pump laser. Next, we scan the pump frequency by modulating the seed laser PZT and observe the Lamb-dip DC signal along with the error signal from $f_{\text{beat}}^{\text{pump}}$ detection, to be sure we lock to the correct comb mode number. Then, we turn the modulation off and engage the lock.

Referencing the pump to the comb allows stabilizing the pump frequency at a controlled offset with respect to the pumped transitions by tuning the $f_{\text{rep}}^{\text{DFG}}$ and $f_{\text{beat}}^{\text{pump}}$. A phase shifter in the RF delay line allows $\pm 500$ kHz adjustments of the $f_{\text{beat}}^{\text{pump}}$ setpoint, and for larger offsets we change $f_{\text{rep}}^{\text{DFG}}$. Another advantage of locking to the comb is that the OODR probe transitions do not have the FM sidebands (as in the Lamb-dip lock case), which simplifies their modeling.

The Allan deviation of the $f_{\text{beat}}^{\text{pump}}$ counted during comb-OODR measurement series #5 (see below) is 12 kHz at 0.7 s. We note that since in the cavity the probe is both co- and counter-propagating with respect to the pump, a shift of the pump frequency causes a split of the probe transition rather than a shift in its center of gravity, and it does not directly translate into uncertainty in the probe line positions.

### 2.2 Comb-OODR: Probing with a frequency comb

The comb probe is an amplified Er:fiber frequency comb (Menlo Systems, FC-1500-250-WG) with $f_{\text{rep}} = 250$ MHz and with a 1.2-m-long polarization-maintaining microstructured silica fiber (PM-MSF) [33] attached at the output. By adjusting the power coupled to the fiber, the center of the probe spectrum can be shifted between 1.6 and 1.8 μm with full-width at -10 dB bandwidth of ~200 cm$^{-1}$. The output of the PM-MSF is connected to a polarization-maintaining fiber-coupled OC and coupled to free space via a collimator (FC1). Two different OCs were used (Advanced Fiber Resources), OC1 designed for the 1.55 μm range and OC2 for the 1.95 μm range (see Table 1).

### 2.2.1 Frequency stabilization and lock to the cavity

The probe comb and the cavity length are stabilized in the same way as in Ref. [20], except that the InGaAs detectors have been replaced by extended-InGaAs detectors to allow operation above 1.7 μm. The $f_{\text{ceo}}$ of the comb is RF-stabilized, while $f_{\text{rep}}$ is stabilized via the lock to the cavity. The comb light is phase-modulated at $f_{\text{PDH}}^{\text{comb}} = 20$ MHz using an electro-optic modulator (EOM) inserted between the oscillator and the amplifier. Comb light reflected from the cavity is collected via the OC, coupled to free space using a collimator (FC2) and dispersed by a reflection grating. A selected part of the spectrum (referred to as the locking point, see Table 1) is incident on a fast extended-InGaAs detector (PD1, EOT, ET-5000). The detector output is demodulated at $f_{\text{PDH}}^{\text{comb}}$ to obtain the PDH error signal. Feedback is sent to the intracavity EOM and PZT in the comb oscillator to lock the comb modes to the sample cavity. Absolute stabilization of $f_{\text{rep}}$ is obtained by referencing it to a tunable RF source (RF2, Menlo Systems, DDS120 Synthesizer) and sending feedback to the sample cavity PZT to keep its length constant for each $f_{\text{rep}}$ setpoint. The values of $f_{\text{ceo}}$ and $f_{\text{rep}}$ are monitored using a counter (Menlo Systems, FXM50) with 1 s gate time.

For efficient coupling of a frequency comb to an optical cavity, the comb $f_{\text{ceo}}$ must be chosen to match the offset of the cavity modes. In our setup, the optimum $f_{\text{ceo}}$ was around 100 MHz, *i.e.,* close to half of the $f_{\text{rep}}$. To circumvent the problem of separating two closely lying beat notes, we downshifted the $f_{\text{ceo}}$ to 20 MHz, where the bandwidth of the RF filters is narrower, using a tunable frequency source (RF1, Anritsu, MG3692A) and a mixer. Another advantage of downmixing is that $f_{\text{ceo}}$ can be tuned by changing the frequency of RF1, without the need to modify the reference frequency and filters of the phase locked loop, which is fixed at 20 MHz.

## 2.2.2 Spectral acquisition

The comb light transmitted through the cavity is coupled into a polarization-maintaining (PM) fiber using a collimator (FC3) and led to a Fourier transform spectrometer (FTS) with auto-balanced detection [34]. The FTS is the same as used in many of our previous works [17, 19, 20, 35]; the only difference is the addition of a new custom-made auto-balancing detector based on two TEC-cooled extended-InGaAs photodiodes (Hamamatsu, G12182-210K) that operate above 1.7 µm. We use both auto-balanced detectors depending on the probed range, *i.e.,* InGaAs, with higher SNR below 1.7 µm and extended-InGaAs with higher SNR above 1.7 µm, depending on the spectral range probed.

The optical path difference (OPD) is calibrated using a frequency-stabilized HeNe reference laser with $\lambda_{ref} = 632.991$ nm (Sios, SL/02/1, fractional frequency stability of $5 \times 10^{-9}$ over 1 h), whose beam is propagating through the FTS on a path parallel to the comb beam. The reference laser and the comb interferograms are measured simultaneously using a DAQ (National Instruments, PCI-5922) and the comb interferogram is resampled at the zero-crossings and maxima of the reference laser.

We record comb probe spectra at different $f_{rep}$ values, tuned using the RF2 in steps of 2.75 Hz, corresponding to a step of ~2 MHz of the comb modes in the optical domain. At each $f_{rep}$ value, we measure two interferograms, one with the pump excitation and one without. To do that, the pump beam is blocked and unblocked on consecutive FTS scans using a shutter. The nominal resolution of the FTS is matched to the $f_{rep}$, and one interferogram is recorded in 3 s. The total acquisition time of a full scan of 130 $f_{rep}$ values is 13 min, including dead time (8.3%) for resampling the comb interferogram and saving the data. For averaging, multiple $f_{rep}$ scans are made in alternating directions. To achieve comb-mode limited resolution we use the sub-nominal sampling-interleaving method [35, 36] that relies on matching the spectral sampling points to the comb mode frequencies by iterative adjustment of the CW reference laser wavelength $\lambda_{ref}$ in post processing to minimize instrumental line shape (ILS) effects. In each measurement, we find $\lambda_{ref}$ that minimizes the ILS for three to five strongest ladder-type OODR transitions and we take the mean as the optimum for that measurement.

**Table 1. Experimental conditions for the OODR measurements using the frequency comb probe. Column 1: Measurement series. Column 2: Transmitted comb spectrum coverage at –10 dB bandwidth. Column 3: Pound-Drever-Hall locking point. Column 4: Incident pump power at the sample, calculated after the first cavity mirror. Column 5: Percentage of the transmitted pump light on resonance with the pump transition. Column 6: Frequency reference for the pump. Column 7: Sample pressure in the cavity. Column 8: Number of scans that were averaged. Column 9: Circulator and auto-balanced detector used during the measurement.**

| | Comb parameters | | Pump parameters | | | | | |
|---|---|---|---|---|---|---|---|---|
| Series | Coverage [cm⁻¹] | PDH locking point [cm⁻¹] | Inc. power [mW] | On-resonance transmission [%] | Freq. reference | Sample pressure [mTorr] | # scans | Components (circulator, FTS detector) |
| Comb #1 | 5510-5640 | 5590 | 833 | 30 | Lamb dip | 170 | 6 | OC2, e-InGaAs |
| Comb #2 | 5745-5960 | 5830 | 807 | 28 | Lamb dip | 175 | 5 | OC2, e-InGaAs |
| Comb #3 | 5865-6095 | 5995 | 639 | 36 | Lamb dip | 181 | 30 | OC2, InGaAs |
| Comb #4 | 5910-6100 | 6015 | 643 | 98 | Lamb dip | 15 | 10 | OC1, InGaAs |
| Comb #5 | 5885-6050 | 6015 | 647 | 98 | MIR comb | 15 | 9 | OC1, InGaAs |

We made 5 measurements series with three different center wavelengths for the comb probe, and two different sample pressures. The conditions of the measurements are summarized in Table 1. In measurement series #1 to #3 the sample pressure was ~170 mTorr, which yields around 30% pump transmission and maximizes the ladder-type OODR signals [18]. In measurement series #4 to #5 the pressure was set to 15 mTorr to avoid saturation of the $2\nu_3$ Doppler-broadened absorption lines and allow observation of the V-type OODR transition in the $2\nu_3$ R(0) line. In measurement series #5, the $f_{rep}$ scan was programmed to yield sample point spacing of ~1 MHz in ±30 MHz range around the $2n_3$ R(0) V-type transition, and 2 MHz elsewhere, so the number of $f_{rep}$ steps was 160. In all measurements, the relative pump/probe polarization was set to the magic angle of 54.7°, which allows the measurement of the OODR probe transition intrinsic line strength [27].

### 2.3 CW-OODR: Probing with a continuous wave laser

The CW probe laser is an external cavity diode laser (ECDL, Sacher, LION model P-1650, driven by a Pilot PZ 500 controller) tunable between 1.6 – 1.75 μm with an output power of ~17 mW near 1666 nm. It has a specified linewidth of <100/300/1000 kHz for time intervals of 1μs/1ms/1s. Its tuning mirror is controlled by a stepper motor and a PZT, which allows 1 GHz/V tuning with a 2 kHz modulation bandwidth. Fine tuning is realized via injection current, with response bandwidth of DC-10 MHz (3 dB). The laser has an internal optical isolator.

The free-space output of the CW probe laser is coupled into a PM fiber using a fiber collimator (FC4) and then split using a fiber coupler (C1). One output of C1 (40%) is used for frequency referencing to a NIR frequency comb as described in Section 2.3.1 below. The other output of C1 (60%) is sent through a PM fiber to a fiber-coupled EOM, then through one of the OCs described previously, and to free space via the fiber collimator (FC1). The CW probe intensity transmitted by the cavity is on the order of 100 μW, and it is detected by an InGaAs detector (PD4, Thorlabs, PDA20CS-EC, 1.87 MHz bandwidth, 13 V/mW at 1650 nm).

### 2.3.1 Frequency stabilization and lock to the cavity

The CW laser is locked to the cavity using the PDH method. Phase modulation is applied at $f_{PDH}^{CW}$ = 4 MHz using the fiber-coupled EOM. The cavity reflected light is collected via the OC, coupled to free space using the fiber collimator (FC2) and incident on an InGaAs detector (PD2, Thorlabs, PDA10CF). Unlike with the comb, no grating is used in front of the detector. The PDH signal demodulated at $f_{PDH}^{CW}$ is sent to a PI²D servo controller (Vescent, D2-125), which also provides the 4 MHz modulation. The feedback is applied to the current control of the CW probe laser. The feedback parameters of the PI²D controller are adjusted to minimize the peak-to-peak amplitude of the error signal. We estimate the closed-loop lock bandwidth to be around 1 MHz from the RF spectrum of the PDH error signal when the lock is engaged.

For absolute frequency stabilization, the CW probe is referenced to the NIR comb (same as the probe comb described in Section 2.2). The NIR comb $f_{ceo}$ is locked at a fixed frequency of 37 MHz using the same feedback electronics as described in Section 2.2.1, with the downshift frequency kept constant at 17 MHz (AimTTI, TGI4162, output 1). The $f_{rep}$ is locked to a tunable RF source (RF3, AimTTI, TGI4162, output 2) via feedback to the intracavity comb PZT. The values of $f_{ceo}$ and $f_{rep}$ are monitored using two channels of a counter (not shown, Menlo Systems, FXM50) with 1 s gate time. The Allan deviations of $f_{ceo}$ and $f_{rep}$ at 1 s are 1 Hz and 1.2 mHz, respectively, indicating stability of the comb mode frequencies of ~1 kHz.

The light from the CW probe and the NIR comb is combined using a fiber coupler (C2) and detected using a 1 GHz bandwidth fiber-coupled InGaAs detector (PD3, New Focus 1611). The second output of C2 is coupled to the FTS for coarse wavelength calibration. The beat note ($f_{beat}^{CW}$) between the NIR comb and the CW probe is filtered and split. One output of the splitter is sent to the third channel of the counter (Menlo Systems, FXM50) to monitor the beat frequency. The other output of the splitter goes to a phase detector (Analog Devices, EV1HMC3716LP4) referenced to a 61 MHz RF source (Analog Devices, AD9914). The feedback is sent via a PI loop filter ($f_{beat}^{CW}$ PI, New Focus, LB1005) to the sample cavity PZT. Controlling the sample cavity length indirectly changes the CW probe frequency via the PDH lock. The bandwidth of this lock is estimated to be 17 Hz based upon the recovery time of the servo output when a small voltage step is applied to the cavity PZT.

The frequencies of the target transitions are known with sub-MHz accuracy from literature and our own comb-OODR measurements and we use their values to calculate $f_{rep}$ and comb mode number, $n$, that places the CW probe laser frequency at the center of the target transition for a known $f_{beat}^{CW}$. To tune the CW probe to the target transition, we adjust the ECDL grating and PZT actuators while using the FTS to measure the CW probe frequency with 250 MHz resolution. For the fine adjustment of the CW probe frequency, we first lock it to the sample cavity and scan it around the target transition. We observe the signal on the oscilloscope, along with the error signal from the phase detector of $f_{beat}^{CW}$. When the sub-Doppler signal and the error signal align, we turn off the fine sweep and engage the lock of the cavity to the beat. We calculate the CW probe frequency using the values of $f_{rep}$, $f_{ceo}$ and $f_{beat}^{CW}$ measured with the counters as $v_{CW} = n f_{rep} \pm f_{ceo} \pm f_{beat}^{CW}$. To find the signs, we lock $f_{ceo}$ and step $f_{rep}$ to find the sign of $f_{beat}^{CW}$ and then we lock $f_{rep}$ and step $f_{ceo}$ to find the sign of $f_{ceo}$.

The $f_{beat}^{CW}$ beat note observed with a spectrum analyzer has a jitter of <500 kHz in 1.1 ms integration time, unchanged when the lock is engaged. This is due to the linewidth of the modes of the RF-locked comb, which is wider than the lock bandwidth. As such, we have a frequency rather than phase lock of the CW laser to the comb. A further improvement of the spectrometer would be to use the high frequency part of the $f_{beat}^{CW}$ error signal to reduce the frequency noise of the comb. The Allan deviation of the counted $f_{beat}^{CW}$ beat note at 1 s is 1.4 kHz, consistent with the estimated stability of the comb modes.

### 2.3.2 Signal acquisition

To increase the absorption sensitivity, we implemented modulation techniques. For measurements of OODR transitions, we amplitude modulate the pump at 500 Hz using a mechanical chopper, and demodulate the signal from PD4 using a lock-in amplifier (Stanford Research, SR830). For measurements of Lamb dips, in which the pump beam is not used, we implemented wavelength modulation spectroscopy (WMS) by applying a sine wave at $f_{WM} = 1$ kHz to the cavity PZT and demodulating the output of PD4 using the same lock-in amplifier, similar to what was done in Ref. [37]. This modulation frequency is above the 17 Hz bandwidth of the $f_{beat}^{CW}$ lock. The peak-to-peak modulation amplitude was around 220 kHz, estimated from the broadening caused by the modulation, and adjusted to maximize the signal without excessive broadening of the transition. In both cases, the detection phase of the lock-in amplifier is set to maximize (minimize) the $X_{in-phase}$ ($Y_{out-of-phase}$) signals. However, to precisely adjust the detection phase, we record both outputs and correct the detection phase in post-processing by choosing a detection phase that maximizes the in-phase component of the signal. To compare signals from different measurements, we recalculate the lock-in amplifier output to relative intensity change using $\Delta I/I = \left( X_{in-phase} \times \frac{sens.}{full\ span} \times C \right)/V_{DC}$, where $sens.$ is the lock-in sensitivity, $V_{DC}$ is the DC signal on the detector (both values vary with measurement series), $full\ span$ is 10 V and $C = \pi/\sqrt{2}$ is a constant accounting for the rms to peak-to-peak value conversion, and for the fact that the lock-in amplifier singles out the first Fourier component of the signal (assuming a square wave produced by the chopper).

To measure the sub-Doppler transitions, we step the CW laser frequency by changing the $f_{rep}$ set point of the NIR comb (via RF3 in Fig. 2b), which in turn changes the sample cavity length via the $f_{beat}^{CW}$ lock to keep $f_{beat}^{CW}$ at 61 MHz. We acquire the lock-in amplifier signal using a DAQ (National Instruments, NI-9219) at 100 Hz rate for 0.1 s, preceded by a dead time of 0.5 s to assure points are taken only after the $f_{rep}$ stepping is completed, that the locks are not in a transient state, and that reading from the counters is updated. The total acquisition time per data point is thus around 0.6 s. The optical frequency step size is in the range of 45 – 180 kHz, depending on the width of the targeted transition. For averaging, multiple frequency scans are made in alternating directions. We recorded selected Lamb-dip, V-type and ladder-type transitions. The Lamb-dip and V-type signals (see Section 3.1.1) were recorded at pressures in the 5-11 mTorr range. For the V-type measurements, the incident pump power at the sample (calculated after the first mirror) was of the order of 40 mW to keep the power broadening low. The relative pump/probe polarization was set to parallel, as it maximizes the R(0) probe signals when an R(0) line is pumped [27]. The weak ladder-type transition (see Section 3.3.1) was recorded at 60 mTorr with 275 mW of pump power to increase the SNR and at both parallel and perpendicular pump/probe polarization to allow rotational assignment.

## 3. Results

### 3.1 Continuous-wave OODR

#### 3.1.1 Frequency accuracy: measurements of the $2\nu_3$ R(0) transition

To estimate the frequency accuracy of our CW-OODR system, we performed measurements of the Lamb dip and the V-type feature in the $2\nu_3$ R(0) transition, with the pump stabilized by referencing to the MIR comb.

Fig. 4 shows the $2\nu_3$ R(0) Lamb dip recorded using the CW probe at 12 mTorr with 45 kHz step size (26 averages). The red curve shows a fit of a model of the first-harmonic WMS signal, based on a derivative of a single Lorentzian line shape. The fit also includes a baseline composed of an offset and a linear slope. The fitted Lorentzian parameters are the peak intensity, width and center frequency. The structure in the residual of the fits is caused by the inaccuracy in the derivative model, as well as possible nonlinearity in the cavity PZT modulation. However, as the structure has odd symmetry, it does not affect the retrieval of the center frequency. In total, we performed 73 scans of the Lamb dip, 26 at 12 mTorr and 47 at 6 mTorr. At these pressures, the mean half-width at half maximum (HWHM) Lorentzian width was found to be 609(6) kHz and 554(4) kHz, respectively. Table 2 reports the mean center frequency obtained from fits to the 73 individual scans corrected for the second order Doppler shift of -308 Hz [26]. The statistical uncertainty of 1.8 kHz is the standard error of the mean (standard deviation divided by the square root of the number of scans). The result agrees within 1.3 kHz with the result obtained using saturated CRDS by Votava et al. [26], which has an uncertainty of 0.23 kHz. For completeness, we also show the reference values from saturation spectroscopy by Ishibashi et al. [38] and Doppler-broadened dual-comb spectroscopy by Zolot et al. [39], which have uncertainties of the order of few hundred kHz.

It is important to note that the result of Votava et al. is corrected by the same second order Doppler shift as well as by a pressure shift. The latter they estimated to be 1.2 kHz at 3.7 μbar, based on the pressure self-shift coefficient reported by Lyulin et al. [40], corrected by a factor of 0.64, which they found by comparing the pressure shift of the R(9) line measured by Lyulin et al. in air for Doppler-broadened lines [41], and measured by them in argon using Lamb-dip spectroscopy. At our measurement pressures of 6 and 12 mTorr (7.9 and 15.8 μbar), the relative pressure shift would be 2.6 kHz, and the precision of our measurement does not allow observing it. We believe this pressure shift might be overestimated, because Barger and Hall [42] reported a pressure self-shift of 75±150 Hz/mTorr for the $\nu_3$ P(7) sub-Doppler line at 3.39 μm, a factor of 5.7 lower than 430 Hz/mTorr used by Votava et al. [26]. Bagaev, Baklanov, and Chebotaev [43] reported, for the same transition, an experimental pressure shift of less than 10 Hz/mTorr and gave a theoretical value of 1 Hz/mTorr near 1 mTorr. Tan et al. [37] also observed anomalously low pressure shift for a Lamb dip in $CO_2$, about 1/5 of the high-pressure coefficient determined from Doppler-broadened spectra. Therefore, we have chosen not to correct our results by the pressure shift, but leave the result from Votava et al. as published by them. However, even considering the different possible values of the pressure shift, our result agrees with Ref. [26] on a 2σ level, demonstrating that our spectrometer is able to measure transitions with an absolute accuracy of ~2 kHz (1 part in $10^{11}$ fractional accuracy). Measurements with higher precision are needed to resolve the issue of the pressure shift.

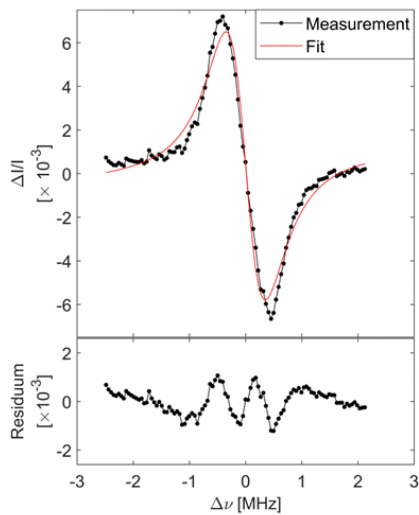

Fig. 4. The WMS Lamb-dip signal in the $2\nu_3$ R(0) transition measured using the CW probe at 12 mTorr (markers, 26 averages, 45 kHz step size), together with a fitted model of the signal (solid, red curve) and residuals in the lower panel.

Measurements of the $2\nu_3$ R(0) V-type CW-OODR transition were carried out at 11 mTorr with 41 mW of pump power (at sample after the first mirror). Fig. 5 shows the V-type when the pump is a) on resonance with the $\nu_3$ R(0) transition and b) offset from it by 3 MHz. In the latter case, the transition is split into two peaks: one broader for the co-propagating pump/probe interaction and one narrower for the counter-propagating. The red curves show fits of models consisting of a single Lorentzian function in a) and a sum of two Lorentzian functions in b). We fit independent peak intensities, widths and center frequencies for the Lorentzian functions, together with a baseline consisting of an offset and a linear slope. The theoretical ratio between the peak values of the counter- and co-propagating features of the split lines is 0.6 (3:5) and the measured ratio for the split lines in Fig. 5b) is 0.629(3). The fitted HWHM widths of the split lines are 3.324(5) MHz and 1.869(3) MHz, so their ratio is 1.779(6), close to the expected value of 1.666 (5:3). For the split lines, we include additional two Lorentzian peaks to model the structure appearing at the center of the transition. We attribute these to the AC Stark shift of the Lamb dip caused by the near-resonant pump field; lock-in demodulation produces the difference in the Lamb dip with and without pump field. When the pump is on resonance and the lines overlap, Fig. 5a), we were not able to model the structure at the center. We speculate that the sharp dip near the center of the line is a cross-over resonance between the two components of the Autler-Townes (AC-Stark) probe absorption peaks that are formed in the probe spectrum by the strong pump field [44].

Table 2 reports the mean center frequencies found from fits to 13 scans of the overlapping and 8 scans of the split V-type transition, where the uncertainty of 4 kHz and 7 kHz, respectively, is the standard error of the mean. The value retrieved from the split V-type transition is more accurate, agreeing within 1σ with the result of Votava *et al.* [26] and with our own Lamb-dip measurement. We attribute the worse agreement of the result from the overlapping V-type transitions in Fig. 5a) to the higher sensitivity of the single-Lorentzian model to the presence of the center feature, which introduces a systematic offset.

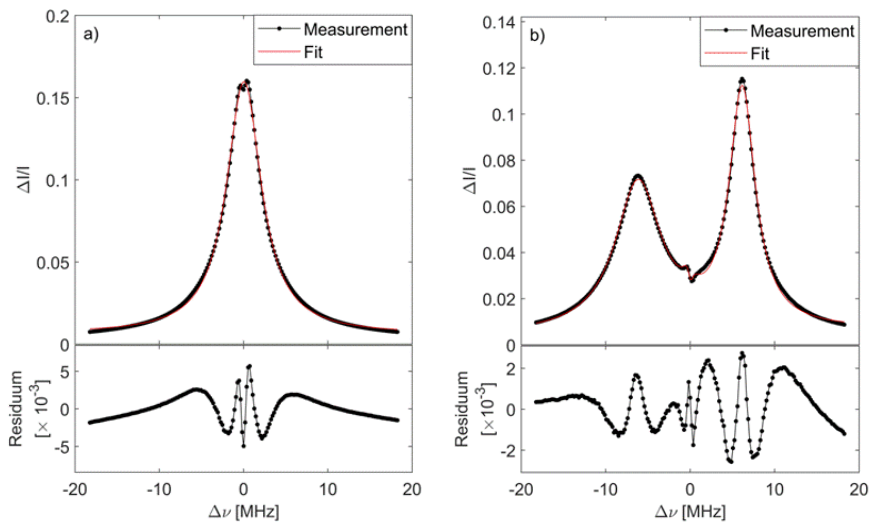

Fig. 5. V-type OODR feature in the $2\nu_3$ R(0) transition measured at 11 mTorr using the CW probe with the pump a) on resonance with the $\nu_3$ R(0) transition (13 averages, 180 kHz step size) and b) detuned by 3 MHz from the $\nu_3$ R(0) transition (8 averages, 180 kHz step size). The red curves show fits of models described in text, and the residuals of the fits are shown in the lower panels.

**Table 2. Center frequency of the R(0) line in the $2\nu_3$ band obtained from the Lamb-dip measurement and V-type measurements using the CW probe and the V-type measurement using the comb probe, compared to reference values from references [26, 38, 39].**

| Measurement series | This work [MHz] | Ref. [26] [MHz] | Ref. [38] [MHz] | Ref. [39] [MHz] |
|---|---|---|---|---|
| CW Lamb dip | 180345064.5287(18) | | | |
| CW V-type (on resonance) | 180345064.566(4) | 180345064.52750(23) | 180345065.08(37) | 180345064.6(2) |
| CW V-type (3 MHz offset) | 180345064.532(7) | | | |
| Comb V-type | 180345064.40(12) | | | |

### 3.1.2 Absorption sensitivity

We estimate the sensitivity of the CW spectrometer for Lamb-dip and OODR detection from the noise in the residuals of the fits shown in Fig. 4 and Fig. 5a). To remove the structure in the residuals, we fitted and removed a 2nd order polynomial from 26 data points at the edge of the fit window. The standard deviation of these data points after baseline removal is $\sigma_{CW} = 9.8 \times 10^{-5}$ and $1.2 \times 10^{-5}$ for the Lamb dip and the V-type, respectively. This gives a minimum detectable absorption coefficient $\alpha_{min} = \sigma_{CW}/L_{eff}$ of $2.2 \times 10^{-9}$ cm$^{-1}$ in $\tau = 15.6$ s and $2.7 \times 10^{-10}$ cm$^{-1}$ in $\tau = 7.8$ s for the Lamb dip and the V-type, respectively. Here, $\tau$ is the acquisition time of one point in the spectrum, multiplied by the number of averages, and $L_{eff} = 2FL/\pi$, where $F = 1180$ is the cavity finesse at 6015 cm$^{-1}$ (see Section 3.2.1) and $L = 60$ cm is the cavity length. The noise equivalent absorption sensitivity NEAS $= \alpha_{min}\tau^{1/2}$, is thus $8.6 \times 10^{-9}$ cm$^{-1}$ Hz$^{-1/2}$ for WMS Lamb-dip detection, and $7.4 \times 10^{-10}$ cm$^{-1}$ Hz$^{-1/2}$ for OODR detection with amplitude modulated pump. We attribute the worse detection limit of the Lamb-dip detection to residual amplitude modulation.

For OODR detection, we also measured the fractional amplitude noise in the transmitted CW beam directly at the output of the lock-in amplifier with the pump beam blocked or having the pump on but far from resonance with a transition. This gave fractional noise of $1.8 \times 10^{-5}$ with a time between the points of 0.6 s, or $1.4 \times 10^{-5}$ Hz$^{-1/2}$, which is a factor of 250 above the shot noise limit calculated for the 0.1 mW power incident on the detector. The corresponding NEAS is $3.1 \times 10^{-10}$ cm$^{-1}$ Hz$^{-1/2}$, a factor of 2 better than estimated from the noise in the residuals.

Neither Karhu *et al*. [7] nor Tan *et al*. [8] stated the sensitivity of their MIR-NIR OODR spectrometers. Karhu *et al*. stated that the SNR for a ladder-type line with $1 \times 10^{-6}$ cm$^{-1}$ peak absorption was 1000. This indicates that the sensitivity was of the order of $10^{-9}$ cm$^{-1}$ for acquisition time of 2.5 s (50 averaged ringdowns with 20 Hz sweep), slightly worse than reported by us, despite their use of a cavity with ~150 times higher finesse. It is impossible to estimate the sensitivity in the work of Tan *et al*. – the residuals of fits have structure on the order of $10^{-9}$ cm$^{-1}$, but the measurement time is not stated.

### 3.1.3   Accurate positions of 2$\nu_3$ R-branch transitions

From the data presented in Section 3.1.1, we conclude that the Lamb-dip measurement provides better center frequency accuracy than the V-type measurement, since the signal is narrower (not power broadened by the pump) and has a simpler line shape, consisting of only one peak. Therefore, we performed Lamb-dip measurements of the R(0) – R(3) transitions in the 2$\nu_3$ band. The R(0) – R(2) lines have previously been measured by Votava *et al.* [26] with sub-kHz accuracy, while the accuracy of their R(3) lines was not specified because of problems with the comb frequency locking.

Table 3Table 3 shows the mean center frequencies of these transitions obtained from our measurements (corrected by the second order Doppler shift), compared to results from Votava *et al.* (measured at pressure of 3.7 µbar, 2.8 mTorr) [26]. The table also specifies the pressure at which our measurement was done and the number of scans included in the determination of the center frequency. The uncertainties are again standard errors of the mean of fits to individual scans. The uncertainty of the R(3) lines is higher because of higher absorption loss and the fewer number of scans measured. Fig. 6 shows the comparison of our results with those from Votava *et al.* [26] where the error bars are combined errors. For the R(0), R(1), R(2, $F_2$) lines, our results agree with 1$\sigma$ with Votava *et al.* [26], and for the R(2, $E$) line the agreement is on a 2.2$\sigma$ level. For the R(3) lines, we assume that our values are more accurate than in Ref. [26], which lacks uncertainties for these lines.

**Table 3. Center frequency of the transitions in the R branch of the 2$\nu_3$ band determined through CW Lamb-dip spectroscopy, compared to reference values from Ref. [26]. Columns 2 and 3 show the pressure at which the measurement was taken, and the number of scans, respectively.**
**\*For $J$ =3, the uncertainties are not reported in Ref. [26].**

| Transition | Pressure [mTorr] | # scans | This work [MHz] | Ref. [26] [MHz] |
|---|---|---|---|---|
| 2$\nu_3$ R(0, $A_1$) | 12 \| 6 | 26 \|47 | 180345064.5287(18) | 180345064.52750(23) |
| 2$\nu_3$ R(1, $F_1$) | 5 | 53 | 180661736.3190(11) | 180661736.31890(44) |
| 2$\nu_3$ R(2, $F_2$) | 5 | 65 | 180974442.1366(4) | 180974442.13610(32) |
| 2$\nu_3$ R(2, $E$) | 8 | 102 | 180974330.4369(3) | 180974330.43790(33) |
| 2$\nu_3$ R(3, $A_2$) | 5 | 20 | 181283407.456(3) | 181283407.4691* |
| 2$\nu_3$ R(3, $F_1$) | 11 | 13 | 181282762.617(8) | 181282762.6304* |
| 2$\nu_3$ R(3, $F_2$) | 11 | 30 | 181283051.578(6) | 181283051.5950* |

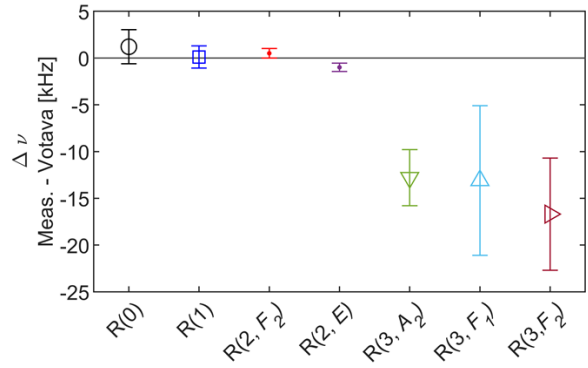

Fig. 6. Center frequencies of the CW Lamb-dip measurements compared to Ref. [26].

### 3.2 Comb-based OODR

#### 3.2.1 Spectral coverage and cavity enhancement factor

Fig. 7a) shows comb spectra transmitted through the cavity at three different center frequencies from measurement series #1-3 (average of all available spectra at the first $f_{rep}$ step with the pump off, see Table 1), demonstrating the wide tunability of the comb-OODR spectrometer. The spectral envelope is a combination of the soliton envelope and the dispersion of the cavity, and the simultaneous spectral coverage is around 200 cm$^{-1}$. Etalon signals induced by various components in the system are also visible. The strong absorption features are the Doppler-broadened CH$_4$ lines in the $2\nu_3$ region, as well as, for measurement series #1, water lines from the ambient air in the free-space path.

We used the Doppler-broadened CH$_4$ lines of the $2\nu_3$ region to evaluate the cavity enhancement factor to be used when fitting the OODR transitions. To do that, we first removed the spectral envelope of the transmission spectra (with pump on) in the cepstral domain [45], *i.e.,* the inverse Fourier transform of the natural logarithm of the spectrum, as described in Appendix A of Ref. [20], using HITRAN2020 [46] parameters for the Doppler-broadened CH$_4$ spectrum. For comb measurement #1, we also included the spectrum of H$_2$O at atmospheric pressure in the baseline, simulated using HITRAN2020 parameters [46]. We interleaved the baseline-corrected transmission spectra to 20 MHz sampling point spacing, sufficient to resolve the Doppler-broadened lines. We divided the spectra into 3 cm$^{-1}$-wide segments and to each segment we fitted a model of the cavity transmission [47] that includes the influence of the phase offset between the comb lines and cavity modes. The fitting parameters were the reflectivity of the cavity mirrors, the comb-cavity phase offset, and a 3$^{rd}$ order polynomial for the remaining baseline. We excluded segments in which transmission of the Doppler-broadened transitions was <10% (strong transition) or SNR was <15 (weak transitions).

Fig. 7b) and c) show the cavity enhancement factors and the comb-cavity phase offsets retrieved from comb measurement series # 1-3 (dots) with corresponding 3$^{rd}$ order polynomial fits (solid lines). The enhancement factors were calculated as $K = \pi/(1\text{-R})$, where R is the reflectivity of the mirrors retrieved from the fit of the model. Since these cavity enhancement factors drop with increasing comb-cavity phase, they are not equivalent to true empty cavity finesse. We suspect that the reason for this discrepancy could be frequency jitter of the comb modes, leading to a stronger reduction of the enhancement for comb modes that are away from resonance with their respective cavity modes. The polynomial fits to the experimentally determined values were used to calculate the enhancement factors at the position of the OODR transitions and fed as fixed parameters to the line fits. The 1$\sigma$ confidence intervals (not shown) of the fits provided the uncertainty contributions to the fitted line intensities.

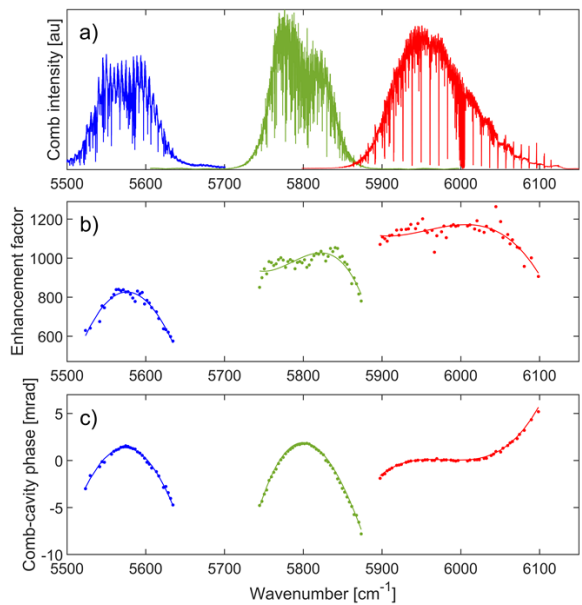

Fig. 7. a) Frequency comb probe spectra transmitted through the cavity showing the spectral coverage of measurements #1-3. b) The cavity enhancement factors (dots) and c) the comb-cavity phase offset (dots) retrieved from fitting the Doppler-broadened lines in measurements #1-3 and the 3rd order polynomial fits (solid curves).

### 3.2.2 Fitting of ladder-type transitions

To detect OODR transitions, we normalized the transmission spectra acquired at each $f_{rep}$ step with pump on to the corresponding background spectra taken with pump off. This operation largely cancels the Doppler-broadened $CH_4$ transitions and the spectral envelope, leaving only the OODR ladder- and V-type transitions, pointing in opposite directions. To remove any baseline structure remaining after normalization, we masked the sub-Doppler OODR features by taking a moving average of the spectrum over windows of >150 GHz and removing spectral points that deviated from the moving mean by a suitable number of standard deviations (typically 0.5-2) of the noise. The remaining spectral structure was modeled as a 5th order polynomial and a series of sine terms. Finally, we divided the normalized spectra by the fitted baseline and interleaved the baseline-corrected spectra at each $f_{rep}$ step, yielding 2 MHz sample point spacing.

The OODR transitions were detected using the routine described in Ref. [20] (Fig. 12 in Appendix C), and the line candidates were inspected by eye to exclude false detections. A few candidates were excluded later during the line fitting stage due to the fit being of poor quality or yielding anomalous line parameters.

To retrieve line parameters of the ladder-type transitions, we applied line fits in windows of ±450 MHz around the center using the same fitting routine and procedures as described in Ref. [20]. We modeled the ladder-type transition as a sum of a narrow Lorentzian line shape for the sub-Doppler peak and a broader Gaussian component induced by velocity redistribution of the pumped population through elastic collisions. The normalization to the Doppler-broadened background was also included in the model, based on parameters from the HITRAN2020 database. The cavity enhancement factor was fixed to the value obtained in Section 3.2.1, and the free parameters of the fit were the integrated absorptions and widths of the Lorentzian and the Gaussian line shape components, a common center frequency of these, and the comb-cavity phase offset. For lines where the SNR of the Gaussian component was below 5, the Gaussian width and the ratio of the Gaussian and Lorentzian integrated absorption were fixed to the mean values obtained for the lines above the SNR threshold in each measurement. In a few cases, the fit window size was reduced to avoid baseline problems on the wings of a line. For the strongest ladder-type lines in measurement series #3, sidebands originating from the phase-modulation of the pump were visible around ±55 MHz from the line center. These were included in the fit models as Lorentzian shapes with their width and intensity fixed to values determined for the strongest line. Fig. 8 shows a plot of the P(1, $A_2$) line at 5913.26 cm$^{-1}$ in measurement #3 (black), together with the fit (red) and its Lorentzian (green) and Gaussian (blue) components. The lower panel shows the residuals, demonstrating good agreement between the model and the experiment.

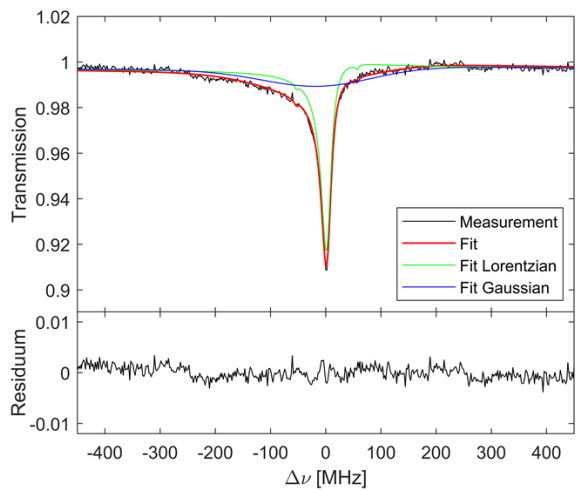

Fig. 8. The P(1, $A_2$) ladder-type transition (black) at 5913.26 cm$^{-1}$ from measurement series #3 and the fit (red). Also shown are the Lorentzian (green) and Gaussian (blue) components of the line shape model. The lower panel shows the fit residuals.

### 3.2.3   Frequency uncertainty

The uncertainties in retrieved OODR line center frequencies have three sources: the line fit, the optimization of the wavelength $\lambda_{\text{ref}}$ of the CW reference laser used for OPD calibration in the FTS, and the pressure shift. The fit uncertainties varied between 50 kHz and 2 MHz depending on the SNR of the transition. To evaluate the uncertainty contribution of $\lambda_{\text{ref}}$ optimization, we fitted three to four selected strong ladder-type transitions in each spectrum analyzed using different $\lambda_{\text{ref}}$ values around the optimum, and we quantified the line center shifts with respect to changes in $\lambda_{\text{ref}}$. Taking the uncertainty of the optimum $\lambda_{\text{ref}}$ to be the standard deviation of the values for the ladder-type lines used for the optimization, we estimated contributions of 50 - 150 kHz to the line center uncertainty, depending on the measurement. Based on the self-induced pressure shift coefficient reported by Lyulin *et al.* [40] we calculated a pressure shift of 120 kHz at 180 mTorr and 10 kHz at 15 mTorr, and we include it in the uncertainty budget. As discussed in Section 3.1.1, this pressure shift is probably overestimated, but we conservatively included it as we did in our previous works [19, 20]. All three contributions were added in quadrature yielding uncertainties of 140 kHz – 2 MHz and 110 – 900 kHz for ladder-type and V-type transitions respectively.

### 3.2.4   Absorption sensitivity

We estimate the sensitivity of comb-based OODR from the noise in the fit residuals. For measurement series #3, the standard deviation of a flat part of the fit residuals shown in Fig. 8 (between -400 MHz to -300 MHz) is $\sigma_{\text{comb}} = 9 \times 10^{-4}$. This yields $\alpha_{\text{min}} = \sigma_{\text{comb}} / L_{\text{eff}}$ of $2 \times 10^{-8}$

cm$^{-1}$ in measurement time $\tau$ = 6.7 h, where the effective length $L_{eff} = 2KL/\pi$ is now defined through the cavity enhancement factor $K$ (see Fig. 7b), equal to 1170 at 5913 cm$^{-1}$, and $\tau$ is the acquisition time of the entire interleaved and averaged spectrum. The NEAS is thus $3.0 \times 10^{-6}$ cm$^{-1}$ Hz$^{-1/2}$. In general, the NEAS depends on the center wavelength of the comb probe spectrum and changes from $1.5 \times 10^{-5}$ cm$^{-1}$ for measurement series #1 to $8 \times 10^{-7}$ cm$^{-1}$ Hz$^{-1/2}$ for measurement series #4. We attribute the increase of noise for longer wavelengths to increased soliton phase noise, which is converted to amplitude noise in the cavity transmission. The measurement series #5 taken at the same comb probe range as measurement #4, has NEAS of $2.2 \times 10^{-6}$ cm$^{-1}$ Hz$^{-1/2}$, a factor of almost 3 worse. These measurements were taken almost 1 year apart, and the reason for the different NEAS is mainly the different soliton phase noise, which is affected by the seasonal temperature and humidity changes in the lab. In any case, we found that the NEAS is not affected by the way the pump is stabilized.

To account for the large simultaneous bandwidth of the comb measurement, we calculate the figure of merit, FOM = NEAS/$\sqrt{M}$, where M is the number of spectral elements in the spectral range reported in Table 1, ranging from $2 \times 10^6$ to $3.7 \times 10^6$ depending on the measurement series. In the measurement series #1, the FOM is $1.1 \times 10^{-8}$ cm$^{-1}$ Hz$^{-1/2}$ per spectral element, and it improves to $4 \times 10^{-10}$ cm$^{-1}$ Hz$^{-1/2}$ per spectral element for measurement series #4. The latter is a factor of 3 and 7.5 better than reported in our previous works [19, 20].

### 3.2.5 Fitting of V-type transitions

We analyzed the V-type features at the center of their respective Doppler-broadened lines in the baseline-corrected transmission spectra with pump on interleaved to 2 MHz sampling point spacing. For each transition, we first eliminated any neighboring Doppler-broadened lines by dividing a section of ±2.5 GHz around the V-type in question by the cavity transmission function of the Doppler-broadened background (excluding the line displaying the V-type transition) calculated using HITRAN2020 parameters and the enhancement factor and comb-cavity phase offset obtained as described in Section 3.2.1. We then masked a ±60 MHz region around the V-type dip and fitted the target Doppler-broadened line in a ±1 GHz window as a Gaussian function with HWHM fixed to 275 MHz, and the center frequency, integrated absorption and comb-cavity phase offset as free parameters. Next, we divided the spectrum by the fit to cancel the Doppler-broadened line. We fitted the remaining V-type feature using the same model as for the ladder-type lines but with an inverted sign. The background absorption and dispersion in the model were now obtained from the fit to the Doppler-broadened line displaying the V-type and the HITRAN model of the surrounding Doppler-broadened lines. The free parameters were the center frequency, the integrated absorption and width of the Lorentzian component and the comb-cavity phase offset. The width of the Gaussian component and the ratio of the Gaussian and Lorentzian integrated absorption were always fixed to the mean values obtained for the ladder-type transitions in the measurement series.

Fig. 9 shows a V-type feature in the $2\nu_3$ R(0) transition at 6015.66 cm$^{-1}$, same as shown in Fig. 5, from measurement series #4 and #5, where the pump was locked to the center of the $\nu_3$ R(0) transition using the Lamb-dip lock and the MIR comb lock, respectively. The fitted model is shown by the red curve, while the Lorentzian and Gaussian components are shown by the green and blue curves, respectively. In Fig. 9a), the FM sidebands are clearly visible and included in the model with their positions and width fixed to values determined from a fit to the strongest ladder-type transitions in the spectrum. The original Doppler-broadened line is displayed in the insets, and the lower panels show the fit residuals. The center wavenumbers retrieved from the fits to these two signals are 6015.663823(4) cm$^{-1}$ and 6015.663825(8) cm$^{-1}$ for measurements series #4 and #5 respectively. The uncertainty of the result from measurement series #5 is higher because of the higher noise levels in the spectrum, caused by the noisier soliton rather than the different way of pump stabilization, as explained in Section 3.2.4. In measurement series #5 we decreased the sampling point spacing to 1 MHz around the V-type transition to check the influence of the sample point spacing on the accuracy of line center retrieval. We found that using all sampling points around the V-type or excluding either half of them from the fitted data changed the retrieved line center only well within the 1σ fit uncertainty. The very good agreement of these two results shows that the two ways of stabilizing the pump frequency yield similar performance. Referencing to the comb has the following advantages over the Lamb-dip lock: 1) it allows pumping lines for which a Lamb dip is not observed in the reference cell, e.g., weaker lines, 2) it provides flexibility in locking the pump at an offset from the target transition (see Section 3.1.1) and 3) it yields OODR signals without sidebands, reducing the number of fitting parameters.

The center frequency of the $2\nu_3$ R(0) V-type transition from measurement series #4 in MHz is inserted in Table 2, showing a 1σ agreement with the result of the CW Lamb-dip measurement of this transition. In our previous work [19], we reported precision of the ladder-type measurements on the order of 150 kHz, and we lacked absolute frequency reference to determine the measurement accuracy. The comparison to our CW Lamb-dip measurement and the work by Votava *et al.* [26] shown here confirms that the accuracy of comb-based OODR is on the 120 kHz level.

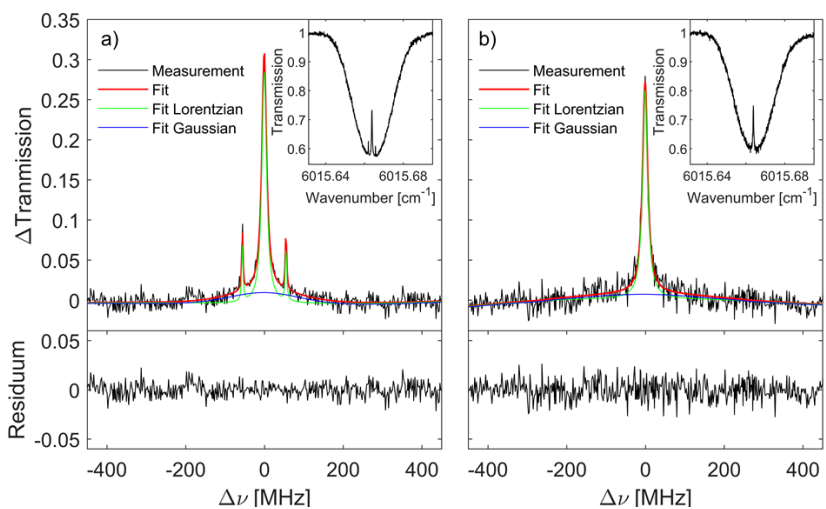

Fig. 9. A V-type feature (black) in the $2\nu_3$ R(0) transition at 6015.66 cm$^{-1}$ using the comb probe with the CW-pump locked to the $\nu_3$ R(0) transition using the a) Lamb-dip lock (measurement #4, 10 averages) and b) MIR comb lock (measurement #5, 9 averages). The red curves show fits of the model, while the green and blue curves show the Lorentzian and Gaussian components of the line shape model, respectively. The insets show the V-type feature on its Doppler-broadened line while the lower panels show the fit residuals.

### 3.3 Comparison to theoretical predictions

#### 3.3.1 Ladder-type transitions

We retrieved line parameters of 37 ladder-type transitions from measurement series #1-3 and report them in Table 4 together with assignments from the effective Hamiltonian [21]. Five strongest transitions in the 5900 – 6050 cm$^{-1}$ range, marked by $^+$ in Table 4, were detected with SNR up to 10 and uncertainties of 1.5 MHz in our initial OODR work using the LN$_2$-cooled single-pass cell [18]. The accuracy of the wavenumbers retrieved here is one order of magnitude better and the results agree with those measured previously within 1σ, except for the R(1) line at 6046.360217(5) cm$^{-1}$, which agrees within 3σ.

**Table 4. Parameters of ladder-type transitions: transition wavenumbers, integrated absorption coefficients and Lorentzian widths, together with the Hamiltonian predictions (wavenumbers and Einstein A coefficients) and vibrational assignments. A multiplicity index *n*, is included in the vibrational labels to distinguish vibrational sub-states with the same symmetry. The lower state of all transitions is $J = (1,A_2)$ with term value 3028.752260 cm⁻¹.**
***Line measured with comb-OODR and CW-OODR (see Fig. 11).***
**⁺Lines reported previously in Ref. [18].**

| Transition wavenumber [cm⁻¹] | Integrated absorption coefficient [10⁻⁹ cm⁻²] | Lorentzian width [MHz] | Hamiltonian transition wavenumber [cm⁻¹] | Hamiltonian Einstein A-coefficient [s⁻¹] | Final vibrational state | Final state J-number | Final state counting number |
|---|---|---|---|---|---|---|---|
| 5553.47949(1) | 5.2(2) | 9.5(3) | 5553.1886 | 0.068028836 | 0 0 2 2 2F₂ | 2A₁ | 75 |
| 5554.08449(2) | 4.8(2) | 10.2(4) | 5554.03274 | 0.049816748 | 0 0 2 2 1A₁ | 0A₁ | 22 |
| 5554.76154(5) | 0.97(8) | 10(1) | 5554.41977 | 0.015272571 | 1 2 0 2 2F₂ | 2A₁ | 76 |
| 5560.35474(4) | 0.54(7) | 7(1) | 5561.13828 | 0.010330305 | 0 0 2 2 1F₁ | 1A₁ | 38 |
| 5567.99856(3) | 1.9(1) | 10.2(9) | 5567.69205 | 0.018179786 | 1 2 0 2 3E | 2A₁ | 78 |
| 5579.81297(4) | 1.5(1) | 10(1) | 5579.85417 | 0.036159297 | 0 0 2 2 2F₁ | 1A₁ | 40 |
| 5582.59579(2) | 3.1(1) | 9.8(5) | 5582.82854 | 0.034408507 | 0 0 2 2 1E | 2A₁ | 80 |
| 5586.02863(4) | 0.5(1) | 5(2) | 5586.60759 | 0.015505148 | 2 1 0 1 1F₁ | 1A₁ | 41 |
| 5592.19447(4) | 1.5(1) | 11(1) | 5592.31434 | 0.011247693 | 0 0 2 2 3F₂ | 2A₁ | 82 |
| 5593.11056(3) | 0.22(7) | 2(1) | 5593.47993 | 0.012171541 | 0 2 1 2 3F₁ | 1A₁ | 42 |
| 5595.46620(3) | 1.6(1) | 9.0(9) | 5595.37193 | 0.02471779 | 0 0 2 2 3E | 2A₁ | 83 |
| 5605.29137(3) | 1.7(1) | 9.0(8) | 5605.67303 | 0.093547725 | 0 0 2 2 3A₁ | 0A₁ | 26 |
| 5613.23494(7) | 1.4(2) | 11(2) | 5612.7252 | 0.008211268 | 0 2 1 2 3E | 2A₁ | 88 |
| 5753.45876(2) | 3.2(2) | 9.3(5) | 5753.4383 | 0.046586508 | 0 1 2 1 3F₂ | 2A₁ | 110 |
| 5762.93157(1) | 3.2(1) | 9.9(3) | 5762.77038 | 0.174820541 | 0 1 2 1 1A₁ | 0A₁ | 32 |
| 5773.19206(2) | 2.1(1) | 8.1(5) | 5773.39084 | 0.074489314 | 0 1 2 1 2F₁ | 1A₁ | 55 |
| 5775.01275(4) | 0.40(4) | 7(1) | 5775.45925 | 0.016202879 | 1 3 0 1 1F₁ | 1A₁ | 56 |
| 5777.885007(8) | 4.8(2) | 9.0(2) | 5777.87683 | 0.070160793 | 0 1 2 1 1F₂ | 2A₁ | 113 |
| 5796.78345(1) | 2.7(1) | 9.3(2) | 5796.62593 | 0.040158333 | 0 1 2 1 2E | 2A₁ | 114 |
| 5800.993131(7) | 7.3(3) | 10.4(2) | 5801.08818 | 0.264815002 | 0 1 2 1 4F₁ | 1A₁ | 58 |
| 5805.37288(1) | 1.64(7) | 9.4(3) | 5805.39456 | 0.02566748 | 0 1 2 1 2E | 2A₁ | 116 |
| 5816.61652(2) | 1.56(7) | 10.7(5) | 5816.94355 | 0.049933036 | 0 3 1 1 2F₁ | 1A₁ | 59 |
| 5828.33573(1) | 1.56(6) | 10.2(3) | 5828.7198 | 0.049099634 | 0 1 2 1 5F₁ | 1A₁ | 60 |
| 5830.15537(2) | 1.10(5) | 9.9(5) | 5830.37531 | 0.013231059 | 0 1 2 1 5F₂ | 2A₁ | 119 |
| *5835.86566(5) | 0.31(3) | 11(2) | 5836.58831 | 0.006396519 | 0 3 1 1 4F₁ | 1A₁ | 61 |
| 5893.86345(3) | 0.14(3) | 4(1) | 5894.16412 | 0.008598897 | 0 3 1 1 4F₂ | 2A₁ | 131 |
| 5897.845489(5) | 3.9(2) | 10.2(1) | 5897.27249 | 0.062246364 | 0 5 0 1 1F₂ | 2A₁ | 132 |
| ⁺5910.806729(5) | 13.4(5) | 10.48(6) | 5910.457 | 0.195235652 | 0 0 3 0 2F₂ | 2A₁ | 133 |
| 5913.263873(6) | 2.37(9) | 10.5(2) | 5914.02737 | 0.205184788 | 0 2 2 0 1A₁ | 0A₁ | 36 |
| ⁺5929.258267(5) | 36(1) | 21.2(1) | 5929.25421 | 1.516920286 | 0 0 3 0 1F₁ | 1A₁ | 67 |
| ⁺5946.587538(5) | 8.1(3) | 13.8(1) | 5946.48342 | 0.687394883 | 1 4 0 0 1A₁ | 0A₁ | 37 |
| ⁺5979.167251(5) | 6.8(2) | 13.3(1) | 5979.16719 | 0.505855445 | 0 2 2 0 1A₁ | 0A₁ | 38 |
| 6009.64830(1) | 0.54(2) | 10.3(2) | 6010.14953 | 0.010814841 | 0 2 2 0 1F₂ | 2A₁ | 139 |
| 6031.16804(4) | 0.14(1) | 10(1) | 6031.57713 | 0.009735941 | 0 2 2 0 2A₁ | 0A₁ | 39 |
| ⁺6046.360217(5) | 17.8(6) | 10.89(8) | 6046.36013 | 0.29950777 | 0 0 3 0 2F₂ | 2A₁ | 141 |
| 6052.70845(4) | 0.41(3) | 12(1) | 6053.145 | 0.005792465 | 0 2 2 0 1E | 2A₁ | 142 |
| 6070.87826(5) | 0.53(6) | 11(2) | 6071.42914 | 0.007093605 | 0 2 2 0 3E | 2A₁ | 143 |

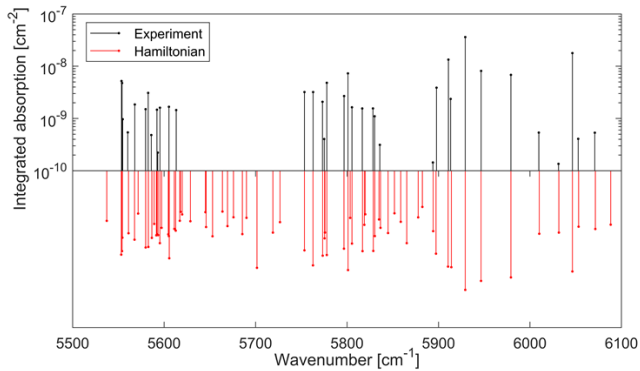

Fig. 10. The measured (black, pointing up) and predicted (red, pointing down) ladder-type transitions, where the predicted Einstein $A$-coefficients have been arbitrarily scaled to match the observed integrated absorption.

The observed ladder-type transitions are plotted in log scale in Fig. 10 (black sticks) together with predictions from the effective Hamiltonian (red sticks, plotted inverted for clarity). The experimental stick height indicates the integrated absorption of the Lorentzian components, while the predicted transitions are represented by their Einstein $A$-coefficients, which have been arbitrarily scaled by a common factor to match the experimental integrated absorption. The observed transitions, with one exception, could be unambiguously assigned using the effective Hamiltonian predictions. In almost all cases, the experimental lines were assigned to the closest predicted transition, though the line at 5605.29 cm⁻¹ was assigned based on the predicted intensity as the closest predicted line appeared a factor of 10 weaker than expected. There was one ambiguous line at 5835.87 cm⁻¹ with two equally plausible assignments, which we resolved by probing it with the CW laser and comparing its intensities measured with parallel and perpendicular pump polarizations. This line is shown in Fig. 11, from the a) comb-OODR measurement series #2, and b) from the CW-OODR measurement. The comb-OODR measurement was done at 175 mTorr with 807 mW of incident pump power, resulting in a HWHM width of 11 MHz, while the CW measurement was done at 60 mTorr with 275 mW of pump power, yielding a narrower line with 3.26(6) MHz linewidth. The fact that the line disappears at parallel pump/probe polarization indicates that it is a Q line, which resolved the Hamiltonian assignment. We note that the comb-OODR measurement took 65 minutes to complete (though it contained many other lines), while the CW measurement took 12 minutes at each pump polarization (3 minutes per scan) and has a much higher SNR. Moreover, the CW data are easier to process, making the CW-OODR a perfect choice for quick evaluation of selected lines. In principle, assignment using polarization-dependent intensity ratios does not require scanning over the line, as was done here for demonstration purposes; it is sufficient to tune the CW probe to resonance, vary the pump polarization, and record the intensity difference. Such measurement can be done in much less than a minute.

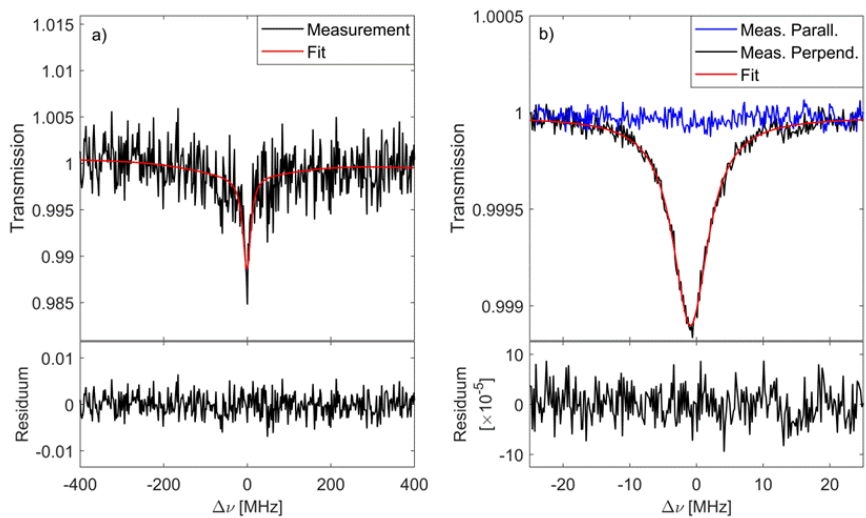

Fig. 11. The ladder-type transition at 5835.87 cm⁻¹ measured a) using the comb probe at 175 mTorr with the relative pump/probe polarization of 57.4⁰ (5 averages, measurement #2) and b) using the CW probe at 60 mTorr and parallel (blue trace) and perpendicular (black trace) pump/probe polarizations (4 averages, 175 kHz point spacing). The red curves show fits to the data.

Assignments to ExoMol [23] and TheoReTS/HITEMP [24] could mostly be performed by comparison of line centers and intensities, though sometimes the assignments to the Hamiltonian served to resolve the choice of assignments in more congested parts of the spectrum. The line at 6031.1 cm⁻¹ remained unassigned to TheoReTS/HITEMP as no plausible transition was found in the predictions. Fig. 12a) shows the discrepancies between the experimental transition wavenumbers and those predicted by the Hamiltonian (green solid squares), ExoMol (blue open circles) and TheoReTS/HITEMP (red stars). The mean and standard deviation of the discrepancies are shown in Table 5. The standard deviations of the wavenumber agreement with the Hamiltonian and TheoReTS/HITEMP are quite similar, but the Hamiltonian predictions have a 3 times lower offset. The ExoMol predictions deviate more from the measurement, particularly towards lower wavenumbers.

The integrated absorption cannot be used for a direct comparison of line intensities between the measurements and predictions, because the population of the pumped level is not known with sufficient accuracy [18]. Therefore, we calculated normalized intensities $I_{N\ obs}$ as the ratios between the integrated absorption of the Lorentzian components of the ladder-type transitions and the V-type transition with the highest SNR in each measurement, similar to what was done in Refs. [18, 20]. We calculated the corresponding predicted normalized intensities as $I_{N\ calc} = (A_{LT}g'_{LT}/v_{LT}^2)/(A_{VT}g'_{VT}/v_{VT}^2)$, from the Einstein $A$-coefficients $A_k$, the upper state degeneracies $g'_k$ and the transition wavenumbers $v_k$ of the ladder-type ($k$=$LT$) and the V-type ($k$=$VT$) transitions, respectively. Fig. 12b) shows the ratios between the experimental and predicted normalized intensities for the Hamiltonian (green solid squares), ExoMol (blue open circles) and TheoReTS/HITEMP (red stars); The $A$-coefficients for the V-type transitions were taken from HITRAN2020. The mean and standard deviations of the ratios of the experimental and predicted normalized intensities are shown in Table 5. The experimental line strengths appear to be smaller than those predicted by all three reference sources by roughly a factor of two. While we at this point do not understand the reason for this discrepancy, we note that similar ratio, 0.38(16), was observed with respect to TheoReTS/HITEMP for the 5 transitions previously observed in Ref. [18].

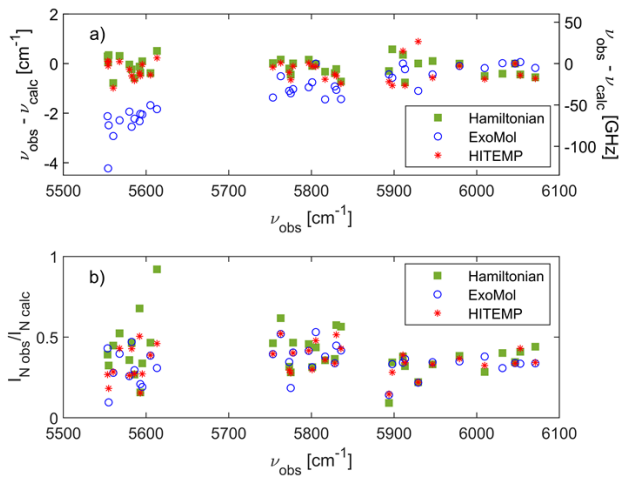

Fig. 12. a) The discrepancies between the experimental ladder-type center frequencies and those predicted by the Hamiltonian (green squares), ExoMol (blue circles) and TheoReTS/HITEMP (red stars). b) The ratios between the experimental normalized intensities and those predicted by the three databases.

**Table 5. The mean and standard deviation of the discrepancies between experimental and predicted ladder-type transition wavenumbers, and the ratios of experimental to predicted normalized transition intensities.**

| Reference | $\Delta v$ [cm$^{-1}$] | | $I_{N\,obs}$ / $I_{N\,calc}$ | |
|---|---|---|---|---|
| | Mean | Std | Mean | Std |
| Hamiltonian | -0.13 | 0.35 | 0.46 | 0.37 |
| ExoMol | -1.2 | 1.0 | 0.41 | 0.34 |
| TheoReTS/HITEMP | -0.30 | 0.41 | 0.40 | 0.31 |

### 3.3.2   V-type transitions

We retrieved parameters of 6 V-type transitions. Assigning them to the Hamiltonian, ExoMol and HITRAN line lists was straightforward, as there was always only one predicted line within a reasonable vicinity that shared the lower state with the $v_3$ R(0) pump transition. The experimental transition wavenumbers together with the assignments and predicted transition wavenumbers and line intensities from the Hamiltonian are given in Table 6. The wavenumber of the $2v_3$ R(0) line, marked with *, is reported from the CW-OODR measurement, while all other lines are from the comb-OODR measurement. Fig. 13 shows a comparison of the experimenal V-type center frequencies to those predicted by the Hamiltonian, ExoMol and HITRAN. There appears to be a systematic shift of the experimental lines from HITRAN, but the agreement is always within the HITRAN uncertainty. The mean and standard deviation of the discrepancies with respect to all line lists are given in Table 7, with ExoMol showing the best agreement.

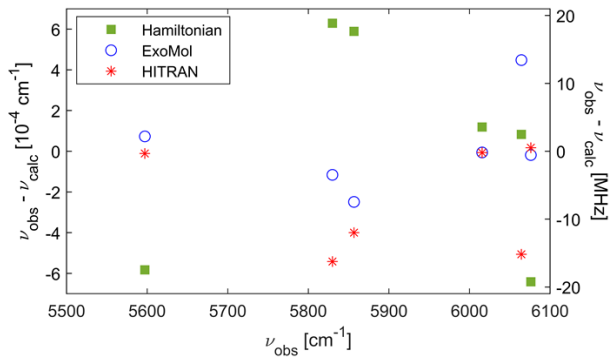

Fig. 13. The discrepancies between the experimental V-type center wavenumbers and those predicted by the Hamiltonian (green squares), ExoMol (blue circles) and HITRAN (red stars).

**Table 6. Experimental and predicted wavenumbers of the V-type transitions, line intensities and assignments from the Hamiltonian.**
**\*Line measured with comb-OODR and CW-OODR, and reported previously in Ref. [18]**

| Transition wavenumber [cm⁻¹] | Hamiltonian transition wavenumber [cm⁻¹] | Hamiltonian line intensity [cm⁻¹/(mol cm⁻²)] | Upper state vibrational assignment | Upper state rotational assignment | Upper state counting number |
|---|---|---|---|---|---|
| 5597.14100(1) | 5597.14158 | $1.32 \times 10^{-22}$ | 0 0 1 2 2F₂ | 1A₂ | 7 |
| 5829.86811(2) | 5829.86748 | $2.74 \times 10^{-22}$ | 0 1 1 1 2F₂ | 1A₂ | 13 |
| 5856.63856(3) | 5856.63797 | $2.64 \times 10^{-23}$ | 0 1 1 1 2F₂ | 1A₂ | 14 |
| 6015.66382729(6)* | 6015.6637 | $5.88 \times 10^{-22}$ | 0 0 2 0 1F₂ | 1A₂ | 18 |
| 6064.377684(9) | 6064.3776 | $3.66 \times 10^{-23}$ | 0 2 1 0 1F₂ | 1A₂ | 19 |

**Table 7. The mean and standard deviation of the discrepancies between experimental and predicted V-type transition frequencies.**

| Reference | $\Delta\nu$ [cm⁻¹] | |
|---|---|---|
| | Mean | Std |
| Hamiltonian | $3.3 \cdot 10^{-5}$ | $5.5 \cdot 10^{-4}$ |
| ExoMol | $2.2 \cdot 10^{-5}$ | $2.4 \cdot 10^{-4}$ |
| HITRAN | $-2.4 \cdot 10^{-4}$ | $2.7 \cdot 10^{-4}$ |

## 4. Conclusions

We developed an OODR spectrometer based on a 3.3 μm CW pump and two cavity-enhanced probes tunable around 1.7 μm: a frequency comb that allows broadband detection of OODR transitions with sub-MHz accuracy, and a CW laser that allows targeting individual transitions with a few kHz accuracy and higher SNR. This combination opens new possibilities in spectroscopy: the comb probe allows surveys of broad spectral ranges in search of previously unknown transitions, which can subsequently be measured with higher SNR and better frequency accuracy using the CW probe. The CW measurements can be done in much shorter acquisition time, and require less post-processing than the comb-based measurements.

The frequencies of both probes are linked to an absolute frequency standard, providing absolute frequency scale. We determined the frequency accuracy of both modes of detection by comparison to previous high-precision measurements of Votava *et al.* [26]. We measured the frequency of the $2\nu_3$ R(0) transition in three different ways: using CW saturation (Lamb-dip) spectroscopy and by probing the OODR V-type feature with the CW laser and with the comb. The Lamb-dip detection yielded 2 kHz accuracy, while OODR detection had 7 kHz and 120 kHz accuracy for the CW and comb probes, respectively. The CW modes of detection yield better accuracy than the comb-OODR because of the better SNR and narrower linewidth of the measured features. For comb-OODR, we use pump power on the order of 800 mW to maximize the transition intensities, which also causes significant power broadening. For CW detection, the Lamb dip is narrower than the V-type feature, because the former is saturated only by the weak probe laser, while the latter is power-broadened by the pump, though less than in the comb-based measurement, due to the use of lower pump power of the order of 40 mW.

Compared to previous MIR-NIR OODR demonstrations, the main novelty of our spectrometer is that it combines two modes of detection and that it probes in the 1.6 –1.8 μm range relevant for methane and other hydrocarbons, rather than in the telecom range. Our CW-OODR spectrometer has similar frequency accuracy and similar or better absorption sensitivity compared to [7, 8], despite the use of a lower finesse cavity. The high sensitivity is obtained by the use of amplitude modulation of the pump and lock-in detection. For the comb-based OODR, we extended the spectral coverage beyond 1.7 μm compared to Refs. [19, 20] and we confirmed frequency accuracy an order of magnitude better than reported previously [17]. Referencing the pump to the MIR comb rather than a Lamb-dip allows pumping transitions that cannot be saturated in a reference cell, and locking the pump at an arbitrary frequency offset with respect to the target transition. Referencing to the comb also provides absolute readout of the pump frequency, without the need to rely on previously published transition frequencies. This will allow pumping transitions whose frequencies are not known with sufficient (kHz-level) accuracy from previous measurements.

We used our spectrometer to measure 37 ladder-type transitions in the $5550 - 6070$ cm$^{-1}$ range from the upper state of the $v_3$ R(0) transition, of which 5 have been detected before [18] with an order of magnitude worse accuracy, and 32 are new. We compare them to theoretical predictions from the new effective Hamiltonian, and the ExoMol and TheoReTS/HITEMP databases, finding best agreement with the Hamiltonian predictions. We detected 6 V-type transitions in the $2v_3$ range, and assigned them using the Hamiltonian, ExoMol and HITRAN line lists, finding best agreement with the ExoMol database. In work to be reported later, we have used the spectrometer for OODR spectroscopy of ethylene, as well as to observe Λ-type OODR transitions and Stark effects in the OODR spectra in methane.

**Funding.** A.F. acknowledges the Knut and Alice Wallenberg Foundation (grant: KAW 2020.0303), and the Swedish Research Council (grant: 2020-00238). K.K.L. acknowledges the U.S. National Science Foundation (grant: CHE-2108458) and the Wenner Gren Foundation (grant: GFOv2024-0010). M.R. Acknowledges the French National Research Agency TEMMEX project (grant: 21-CE30-0053-01).

**Data availability.** Data underlying the results presented in this paper are available from the corresponding author on reasonable request.

**Disclosures.** The authors declare no conflicts of interest.